\documentclass[twoside]{dis08}
\usepackage[latin1]{inputenc}
\usepackage[dvips]{graphicx,epsfig,color}
\usepackage{wrapfig,rotating}
\usepackage{amssymb,amsmath,array}
\usepackage{subfigure}
\begin{document}
\title{New Results from the H1 Collaboration}

\author{Andr\'e Sch\"oning
%
\thanks{On behalf of the H1 collaboration.}
%
\vspace{.3cm}\\
%
University of Zurich - Department of Physics \\
Winterthurerstr. 190, CH-8057 Zürich - Switzerland \\
and \\
DESY \\
Notkestr. 85, D-22607 Hamburg - Germany \\
%
}

\maketitle

\begin{abstract}
After more than 15 years, the operation of the world's unique electron-proton collider
HERA ended in June 2007. 
The data collected by the H1 experiment 
correspond to a total integrated luminosity of about $0.5$~fb$^{-1}$. 
In 2007 dedicated runs were taken at lower proton beam energies to
measure the longitudinal proton structure function $F_L$. 
In this talk new results, many of them exploiting data taken after the
HERA upgrade in the year 2000, and the first measurement of $F_L$ are presented.
\end{abstract}
%

\section{Introduction}
After more than 15 years of successful data taking at HERA the H1
collaboration has started to finalise the data analyses and is entering a new
era of high precision studies of the proton structure.
The basis are the high quality data taken by the H1 experiment in 
electron- or positron-proton collisions at HERA I (year 2000 and before) and
at HERA II (2003-2007) with a total integrated luminosity of about
$0.5$~fb$^{-1}$.
At the upgraded HERA II longitudinally polarised electron or positron beams were provided
regularly with polarisation degrees of 40\% and more.
The data were collected at center of mass energies of about $320$~GeV
($300$~GeV before 1999) and $225$ $(275)$ GeV in dedicated runs with lowered
proton beam energies of $E_p=460$ $(575)$ GeV in order to perform a first direct
measurement of the longitudinal proton structure function $F_L$.

The main topics covered by this talk are deep inelastic scattering at high $Q^2$,
tests of electroweak physics, searches beyond the Standard Model (SM), the
study of the proton structure, jet physics, the measurement of the strong 
coupling and the study of exclusive final states like charm production.

\section{Electroweak Physics and High \boldmath{$Q^2$}}
\begin{wrapfigure}{r}{0.5\columnwidth}
\vspace{-0.4cm}
\centerline{
\vspace{-1.1cm}
\includegraphics[width=0.52\columnwidth]{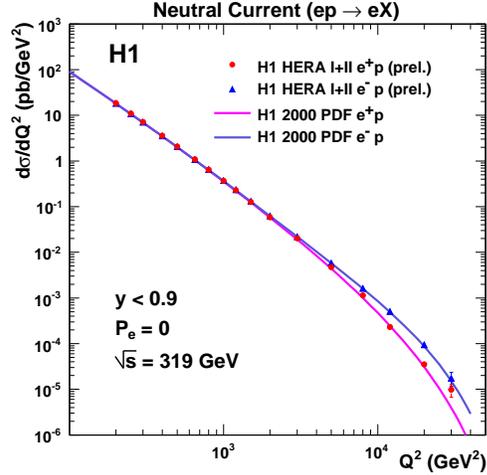}}
\caption{Single differential cross section $d\sigma/dQ^2$ for $e^-p$ and
  $e^+p$ as function of $Q^2$. }\label{fig:ncdis}
\end{wrapfigure}

Using the full statistics of high energy HERA data the single differential
neutral current (NC) cross section in deep inelastic scattering (DIS)
is measured as function of the negative
four momentum transfer squared $Q^2$. 
The cross sections are shown in
Fig.~\ref{fig:ncdis} for $e^+p$ and $e^-p$ collisions separately.
The cross sections, measured over a large $Q^2$ range, change by more than six
orders of magnitude and are well described by the SM prediction based on the proton density
function (PDF) previously published by H1 using only the
HERA I data~\cite{ref:dis_h1pdf}.
For $Q^2 \lesssim 1000$~GeV$^2$ the DIS interaction is dominated by photon
exchange and the cross section scales like $1/Q^4$. For larger values of $Q^2$
the $Z$ exchange and the $\gamma-Z$ interference contribute, the latter being
responsible for the different cross sections between $e^+p$ and $e^-p$ at high
$Q^2$.

\section{Searches}
The existence of new particles or new interactions of fermions or bosons would manifest as
deviations from the SM expectations at high $Q^2$. These deviations can
be described in the context of contact interactions if the mass of the
new particles is much higher than the available 
energy of the studied reactions.
Deviations from the SM are also expected in composite models if quarks or
leptons have a finite size. Using the full HERA I+II data quark radii larger
than $0.74 \cdot 10^{-18}$~m can be excluded.

\begin{wrapfigure}[20]{l}{0.5\columnwidth}
\vspace{-0.43cm}
\centerline{\includegraphics[width=0.52\columnwidth]{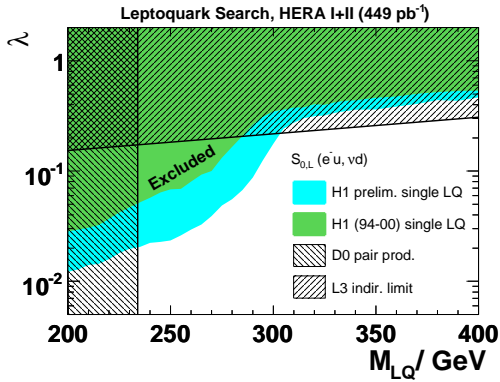}}
\vspace{-0.3cm}
\caption{Exclusion limits at 95\% CL on the coupling $\lambda$ as a function of
  the leptoquark mass for $S_{0,L}$ using the HERA I+II data. The indirect limits of LEP (L3 and
  OPAL) and the direct limit from Tevatron (D0) are shown
for comparison. The published H1 limit from HERA I is also shown.}\label{fig:LQlimit} 
\end{wrapfigure}

New particles like leptoquarks (LQ), which are predicted by Grand Unified Theories, might have masses
as light as the electroweak energy scale and could be resonantly produced at colliders. 
A search for first generation LQs (coupling to first generation
leptons) has been performed by H1 using the full HERA data set. 
No significant deviation from the SM expectation is found and limits on the
Yukawa coupling $\lambda$ as function of the LQ mass are set. The exclusion
regions for the $S_{0,L}$ type LQ, which corresponds also to the scalar
d-quark in supersymmetric models, are shown in Fig.~\ref{fig:LQlimit} and
compared with results from other collider experiments. Similar results are
also obtained in a search for LQs coupling to the first and second lepton generation.

\begin{figure}
\hskip -0.3cm
\subfigure 
{
    \includegraphics[width=0.55\columnwidth]{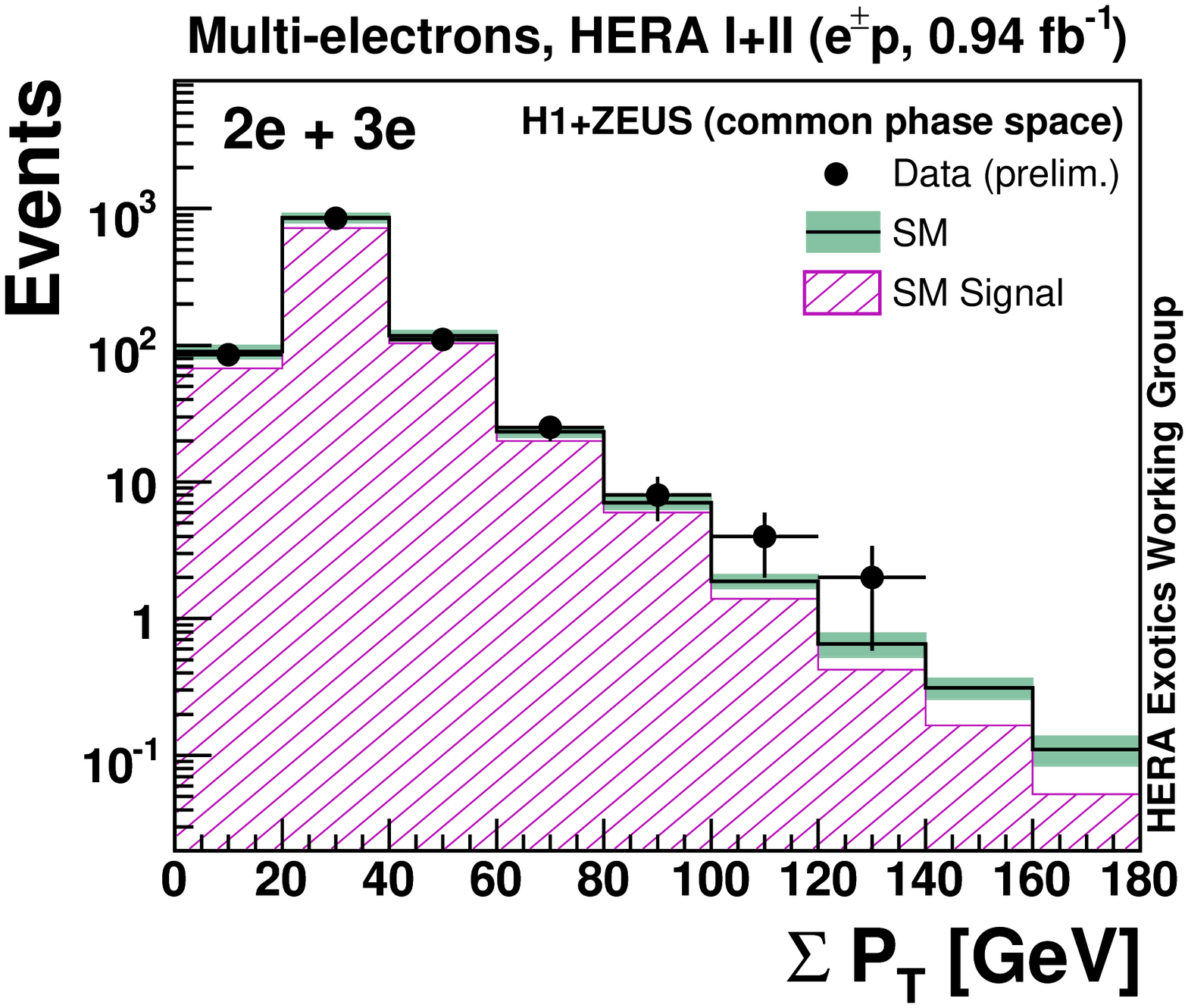}
}
\hskip -0.6cm
\subfigure 
{
    \includegraphics[width=0.52\columnwidth]{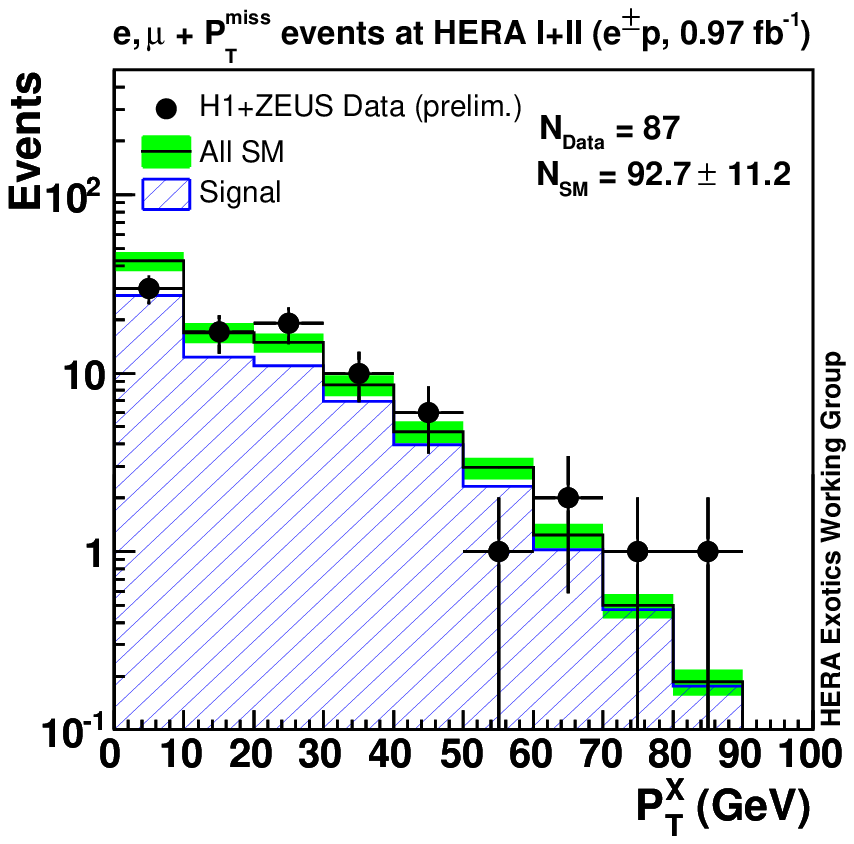}
}
\caption{Left: Distribution of the scalar sum of the transverse momenta of the
  combined 2 and 3 electron events using all HERA I+II data sets from H1
  and ZEUS. The data (points) are compared to the SM expectation (open histogram).
Right: The hadronic transverse momentum distribution of the combined H1 and
ZEUS $e^\pm$ data (HERA I+II). The data (points) are compared to the SM
expectation (open histogram). The signal component of the SM expectation,
dominated by $W$ production, is given by the hatched histogram.
$N_{Data}$ is the total number of events observed, $N_{SM}$ is the total SM
expectation. Both: The total error on the SM expectations is given by the shaded band.}\label{fig:leptons}
\end{figure}

In addition to the previous model dependent searches interesting final
state topologies are investigated to search for possible deviations from the
SM expectation. Of special interest are final states with leptons because of
their clear signature and low background. In
Fig.~\ref{fig:leptons} (left) the scalar sum of the transverse
momentum, $\sum P_T$, of final states containing two or three clearly identified high $P_T$
electrons is shown using all HERA data from H1 and ZEUS combined corresponding
to a total integrated luminosity of ${\cal L}=0.94$~fb$^{-1}$. 
The main SM process is the electron pair production $ep
\rightarrow (e)ee X$ where often one electron undetected escapes down the beampipe. Overall,
good agreement between the data and the SM expectation is found. At
$\sum P_T>100$~GeV nine events are observed in total, compared to a SM
expectation of $5.3 \pm 0.6$ events. The excess of data at
$\sum P_T>100$ comes from the H1 dataset alone \cite{ref:multil} and is not
seen in the ZEUS data. 
Similar analyses containing one or several muons
in the final state have been recently published by H1 using the full HERA data set
\cite{ref:multil}. Also in these channels with muons, which are awaiting
combination with ZEUS data, a slight excess in the region $\sum P_T>100$ is found.

Another interesting final state topology consists of a high $P_T$ lepton and missing
transverse momentum. In the SM this topology is mainly due to the
production of single $W$-bosons with subsequent leptonical decay. Combining the HERA I+II
data sets from H1 and ZEUS, corresponding to an integrated luminosity of 
${\cal L}=0.97$~fb$^{-1}$, events containing either an electron or a muon are
selected. In the common H1/ZEUS phase space in total 87 events are observed
corresponding to a SM expectation of $93\pm 11$ events. 
The measured transverse momentum distribution of the hadronic final state $P_T^X$ is shown
in Fig.~\ref{fig:leptons} (right) and compared to the SM expectation. 
Overall good agreement between data and Monte Carlo is
found. For $P_T^X>25$~GeV, a region where an excess was reported by the H1
collaboration \cite{ref:isollepton}, 29 events are observed in the common
H1/ZEUS phase space in agreement to a SM expectation of 25.3$\pm$3.2 events.

The W enriched isolated lepton sample from H1 is also used to measure the
W-polarisation by studying the decay angle in the rest frame of the W-boson. 
Within the large errors good agreement with the SM expectation is found. 
For more details see \cite{ref:talk_wpola}. 
The polarisation amplitudes depend on the production mechanism of the
W-boson and are SM-like for ordinary single $W$ boson production. However, they
differ if the W-boson originates for example from a decay of a top quark, which can be
anomalously produced at HERA, if the magnetic type coupling
$\kappa_{tu\gamma}$ or the vector type coupling $v_{tuZ}$ are different from
zero. A more direct way of testing these anomalous couplings is the search for
top quark final states. Such a search is performed
by H1 and no sign for anomalous top quark production is found, see also 
\cite{ref:talk_wpola}. A limit of $\kappa_{tu\gamma}<0.14$ is set from the
H1 data alone, which is the most stringent limit from collider experiments.


\section{Proton Structure}
\begin{wrapfigure}{r}{0.6\columnwidth}
\vspace{-1.0cm}
\centerline{\includegraphics[width=0.62\columnwidth]{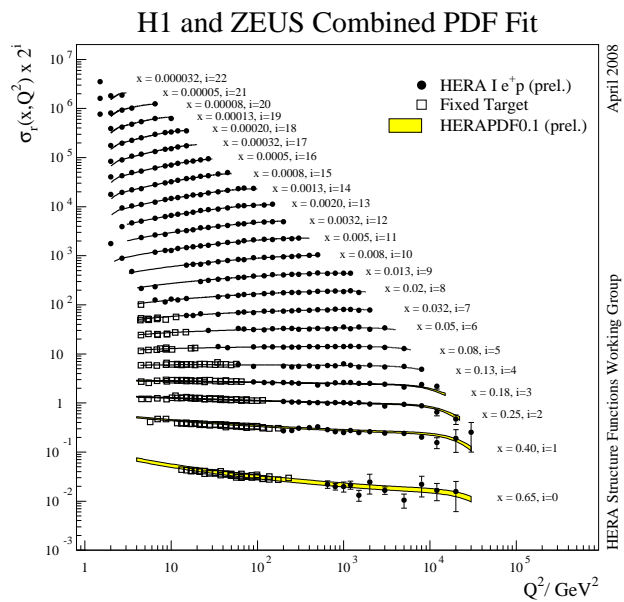}}
\vspace{-0.3cm}
\caption{The reduced cross section $\sigma_r(x,Q^2)$ as function of $Q^2$ for
  various values of $x$ obtained by combining H1 and ZEUS data from HERA~I. 
  Also shown is a new fit of the proton PDFs to the combined data set. 
 }\label{fig:sigmareduced}
\end{wrapfigure}
The most precise determination of the proton structure is achieved by
measuring the inclusive cross section of deep inelastic scattering (DIS).
Double differential cross sections were measured by the H1 \cite{ref:dis_h1, ref:dis_h1pdf} and ZEUS
\cite{ref:dis_zeus} collaborations using the HERA I data  as function of $Q^2$ and the Bjorken variable $x$, which describes
the fractional momentum of the scattered parton in the naive quark parton
model (QPM).
Due to the large centre of mass energy a wide kinematic range can be
explored by HERA to study the QCD dynamics at small $x$.
Results of a model independent combination of the published H1 and
ZEUS measurements based on neutral current (NC) and charged current (CC) data
are presented at this workshop \cite{ref:talk_feltesse}. 
For the combination an averaging procedure is used, which is
based on the assumption that H1 and ZEUS experiments measure the same cross section at the
same kinematic points. This method takes systematic uncertainties both uncorrelated and correlated for
the two experiments into account. The H1-ZEUS combined double differential cross
section results have substantially reduced errors compared to the individual
experiments and represent the most
precise determination of the DIS cross sections at HERA. 
The combined data also serve as basis for a new precise fit of the proton
PDFs, see presentation by \cite{ref:talk_mandy}. The H1 and ZEUS combined
cross sections and the new fit of the proton PDF is shown in Fig.~\ref{fig:sigmareduced}.

For not too high values of $Q^2$, where electroweak effects have to be taken into
account, the reduced NC cross sections can be written as
$\sigma_r = F_2 - \frac{y^2}{Y_{+}} F_L$, where y is the inelasticity and $Y_{+}=1+(1-y)^2$.
The structure function $F_2$, which is dominant except for
very large values of $y$, is related to the charge density of partons in the
proton. The structure function $F_L$, which can be as large as $F_2$,
corresponds to the cross section of longitudinally polarised virtual photons.
The QPM predicts $F_L=0$. In QCD, however, $F_L$ is non-zero and can be related to the
gluon density of the proton. A measurement of $F_L$ can therefore be regarded
as a consistency check of perturbative QCD calculations based
on the gluon density extracted from scaling violation.
\begin{figure}[t]
\hspace{-0.6cm}
\centerline{\includegraphics[width=1.0\columnwidth]{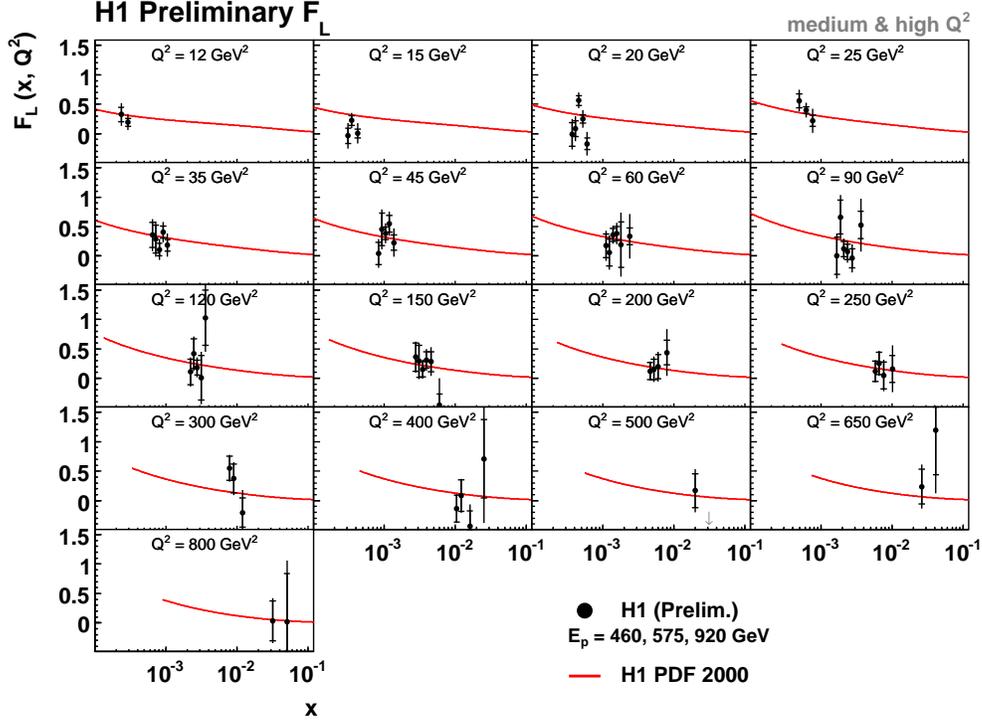}}
\vspace{-1.0cm}
\caption{The longitudinal structure function $F_L(x,Q^2)$. The inner error
  bars denote the statistical error, the full error bars include the
  systematic error. The luminosity uncertainty is not included in the error
  bars. The curve represents the NLO QCD prediction derived from the H1 PDF
  2000 fit to previous H1 data.}\label{fig:FLall}
\end{figure}

Previous determinations of $F_L$ at HERA \cite{ref:dis_h1pdf, ref:dis_h1fl2} were based on extrapolating the
structure function $F_2$ to large values of y (corresponding to small values
of x). These indirect determinations were assuming for $F_2$ a power law function:
$F_2 \sim x^{-\lambda(Q^2)}$.
At this workshop the first direct measurement of $F_L$ is presented, see  for details
\cite{ref:talk_fl1, ref:talk_fl2, ref:talk_flzeus}.
The measurement of $F_L$ requires several sets of cross sections at fixed
$Q^2$ and $x$ but at different $y$. This was achieved by lowering the proton
beam energy according to $y=Q^2/(4xE_pE_e)$ whilst keeping the positron beam
energy $E_e$ constant. The datasets with lowered proton beam energies of
$E_p=460$ ($E_p=575$)~GeV
were taken in dedicated runs from April to June 2008 and correspond to integrated luminosities
of ${\cal L}=12.4$ ($6.2$)~pb$^{-1}$. The big challenge of the $F_L$
measurement is the determination of the cross section at large values of $y$,
which correspond to small scattered positron energies according to $E_e^\prime
\sim (1-y) \; E_e $. The scattered positron is detected either in the SPACAL
detector at large polar angles ($\Theta>155^\circ$) or in the LAr calorimeter
($\Theta<155^\circ$). 
For small positron energies the background
from photoproduction, where one of the hadronic final state particles is wrongly
identified as positron, is large. This background is estimated by 
using Monte Carlo simulations and data by measuring the charge of the assigned track of the scattered
positron candidate, which in photoproduction turns out to be almost equally
populated between electron and positron candidates. 
The estimated photoproduction background is statistically subtracted from the
DIS sample. From the $y$
dependence of the cross section for fixed 
$Q^2$ and $x$ the structure function $F_L$ is measured. The results are
presented in Fig.~\ref{fig:FLall}, where $F_L$ is shown as function of $x$ in
various bins of $Q^2$.
The results are consistent with the predictions obtained from the H1 PDF 2000 fit
\cite{ref:dis_h1pdf}, which was performed using the high proton energy data from HERA~I
only.

\begin{figure}[t]
\centerline{\includegraphics[width=0.75\columnwidth]{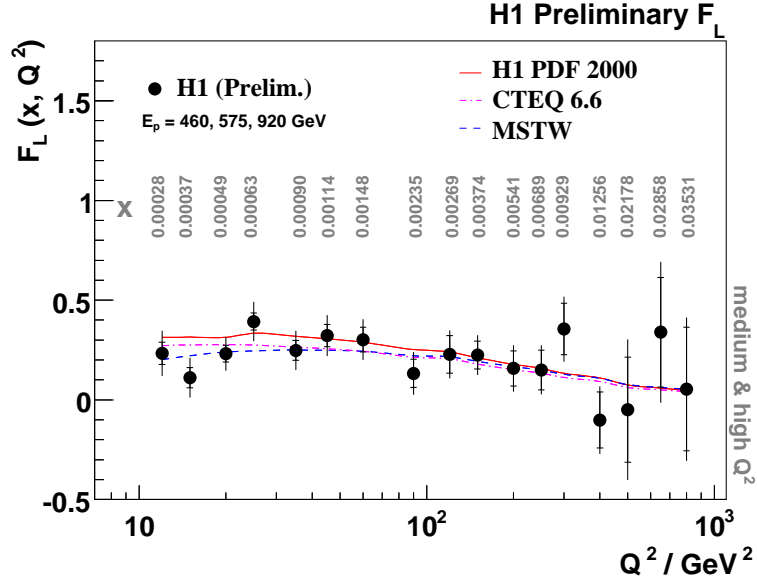}}
\vspace{-0.3cm}
\caption{The longitudinal structure function $F_L$ shown as function of $Q^2$
  at the given values of $x$.     
The inner error bars denote the statistical error, the full
    error bars include the systematic errors. The luminosity uncertainty
    is not included in the error bars.
       The solid  curve describes the expectation
     on $F_L$  from the H1 PDF 2000 fit using NLO QCD.
     The dashed (dashed-dotted) curve is
      the expectation of the MSTW (CTEQ) group using NNLO (NLO) QCD.
      The theory curves connect predictions at the
      given $(x,Q^2)$ values by linear extrapolation.
 }\label{fig:FLaveraged}
\end{figure}

The values on $F_L$ resulting from averages over $x$ at fixed $Q^2$ are shown 
in Fig.~\ref{fig:FLaveraged}. The results are compared with QCD predictions
based on different parametrisations of the proton PDFs
and show good agreement. The results confirm DGLAP QCD predictions for
$F_L(x,Q^2)$, determined from previous HERA measurements, which are
dominated by a large gluon density at low $x$. These results present
 the first measurement of $F_L$ at low $x$. The results
obtained using the SPACAL for positron detection
($Q<90$~GeV$^2$) 
were recently published \cite{ref:flpub}.

\section{Exclusive Final States}
\subsection{Perturbative QCD and Jets}
\begin{wrapfigure}{r}[0cm]{0.55\columnwidth}
\vskip -0.9cm
\centerline{\hskip -8.0cm
\includegraphics
[bbllx=570,bblly=35,bburx=0,bbury=640,width=0.53\columnwidth,angle=90]{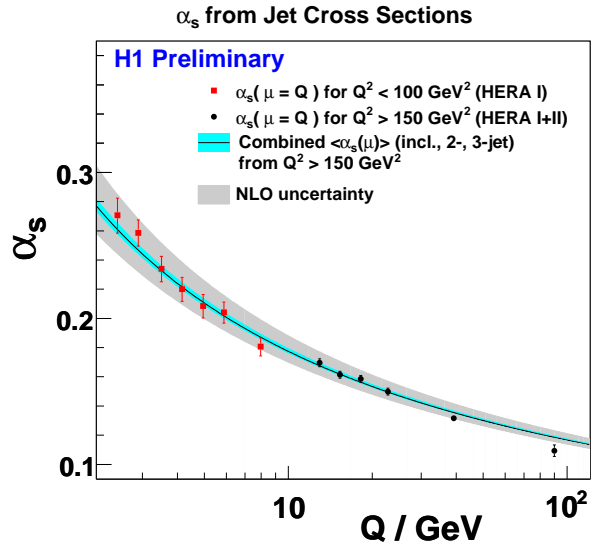}}
\caption{Results of the fitted values of $\alpha_S(\mu=Q)$ from the
low $Q^2$ measurements (red points) and the high $Q^2$ measurements
(black points). The error bars denote the total experimental uncertainty of
each data point. The solid curve shows the result of evolving $\alpha_S(M_Z)$
obtained from a fit to the inclusive, dijet and threejet normalised cross
section at high $Q^2$, with the inner blue band denoting the correlated
experimental uncertainties and the grey band denoting the theoretical
scale uncertainties, PDF uncertainties and the hadronisation corrections.}
\label{fig:alphaS}
\end{wrapfigure}
The production of jets at high transverse momentum can be theoretically well described in
the framework of perturbative QCD (pQCD) calculations. For HERA only NLO
calculations for jet production are available and the largest theoretical
uncertainties come from the unknown normalisation and factorization scale
reflecting our ignorance about higher order contributions.
A pity, as the production of two
or more jets in DIS gives direct access to the strong coupling $\alpha_S(M_Z)$, which
is one of the fundamental parameters in particle physics. 
The strong coupling can experimentally be measured at HERA with high precision due
to the large statistics of the HERA I+II datasets.
The H1 collaboration performed two new measurements of  $\alpha_S$, one in the
low $Q^2$ region, where the scattered electron is detected in the backward
SPACAL and one in the high $Q^2$ region, where the scattered electron is
detected in the LAr calorimeter.

The measurement at low $Q^2$ is based on HERA~I data only \cite{ref:talk_alphas}. The kinematic
region is defined by $5<Q^2<100$~GeV$^2$ and $0.2 <y< 0.7$.
Jets are selected in the Breit frame using the inclusive $k_T$ algorithm with
a minimum transverse momentum of $5$~GeV.
Inclusive jet cross sections are measured differentially as function of $Q^2$
and the jet transverse momentum $E_T$. The measured inclusive cross sections
are compared to NLO QCD calculations using the NLOJET++ \cite{ref:NLOJET} program. The
calculations are performed in the $\overline{\rm MS}$ scheme  for five massive quark
flavors and for the PDF parametrisation CTEQ65 \cite{ref:CTEQ65}.
As factorisation scale $\mu_F=Q$ is used, whereas the normalisation scale
is chosen to be $\sqrt{(Q^2+E_T)/4}$. The 
strong coupling is extracted from the measured distributions and the results are shown as function of the $Q$
in Fig.~\ref{fig:alphaS} (red points). A fit to the low $Q^2$ data yields:
\begin{equation}
\alpha_S(M_Z)=0.1186 \ \pm \ 0.0014 {\rm (exp)} \ ^{+0.0132}_{-0.0101}  {\rm
  (scale)} \ \pm \ 0.0021  {\rm (PDF)} \quad ,
\end{equation}
where the first error is due to experimental uncertainties, the second and
dominating one from the scale uncertainties and the third from the
uncertainty of the PDF. 

The measurement of jet rates at high $Q^2$ is based on HERA~I+II data
corresponding to a total integrated luminosity of ${\cal L}=395$~pb$^{-1}$ \cite{ref:talk_alphas}. 
Jet rates are measured for $160<Q^2<15000$~GeV$^2$ inclusively, and for 2-jet and
3-jet final states. In contrast to the low $Q^2$ analysis relative jet rates are measured
normalised to the inclusive NC cross section. This reduces
the experimental error as the luminosity uncertainty cancels completely 
and the PDF uncertainties to a large extend. The relative jet rates are measured again
as function of $Q^2$ and $E_T$ in the Breit frame. The strong coupling
$\alpha_S$ is extracted in a same way as at low $Q^2$. 
The PDFs of the proton are taken from the CTEQ65M set \cite{ref:CTEQ65M}. 
The extracted $\alpha_S$ values are shown in Fig.~\ref{fig:alphaS} as function
of $Q$. A fit to all the high $Q^2$ jet data yields:
\begin{equation}
\alpha_S(m_Z)=0.1182 \ \pm \ 0.0008 {\rm (exp)} \ ^{+0.0041}_{-0.0031}  {\rm
  (scale)} \ \pm \ 0.0018  {\rm (PDF)} \quad ,
\end{equation}
well compatible with the result obtained at low $Q^2$. This result has a
strikingly small experimental error below 1\% and is dominated by the large scale
uncertainty. The importance 
of calculating the higher order corrections in order to achieve a
precise determination of $\alpha_S$ from the HERA data has to be stressed.

\begin{wrapfigure}[25]{r}{0.5\columnwidth}
\vskip -0.8cm
\centerline{
\includegraphics[width=0.42\columnwidth]{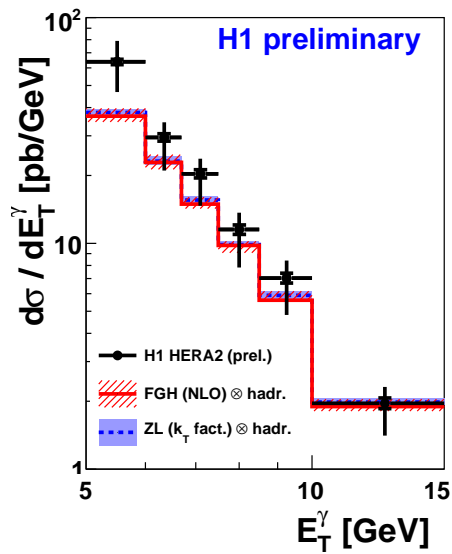}}
\vskip -0.5cm
\caption{Inclusive prompt photon cross section as function of
  $E_T^\gamma$. 
The inner error bars represent the statistical errors, while the outer are the
statistical and systematical errors added in quadrature. The measured cross
section is compared to the Fontannaz-Guillet-Heinrich (continuous red) and
the Zotov-Lipatov (dashed blue) calculations.
}
\label{fig:promptgamma}
\end{wrapfigure}
Hard QCD processes are also studied by H1 in prompt photon photoproduction
using the full HERA II data. 
The cross sections are measured in the visible range defined
  by $5<E_T^\gamma<15$~GeV, $Q^2<1$ GeV$^2$, $0.1 < y_h < 0.7$
\footnote{The variable $y_h$ is defined as $y_h=\sum (E-pz)/2E_e$ where the sum
is executed over all final state particles.} 
and
  $z=E_T^\gamma/E_T^{\textrm{photon-jet}}>0.9$.
The results, see for more details \cite{ref:promptgamma}, indicate that the cross
sections are in good agreement with QCD NLO calculations \cite{ref:FGH} and with QCD LO calculations
based on $k_T$ factorisation \cite{ref:ZL} in exclusive final states where a high $E_T$ jet is
present. The data, however, overshoot the predictions in the inclusive prompt
photon sample for which no jet requirement is  applied, see Fig.~\ref{fig:promptgamma}.

\subsection{Soft QCD and Diffraction}
\begin{figure}[t]
\centerline{
\includegraphics[width=0.52\columnwidth,clip=,bbllx=0,bblly=0,bburx=375,bbury=295]{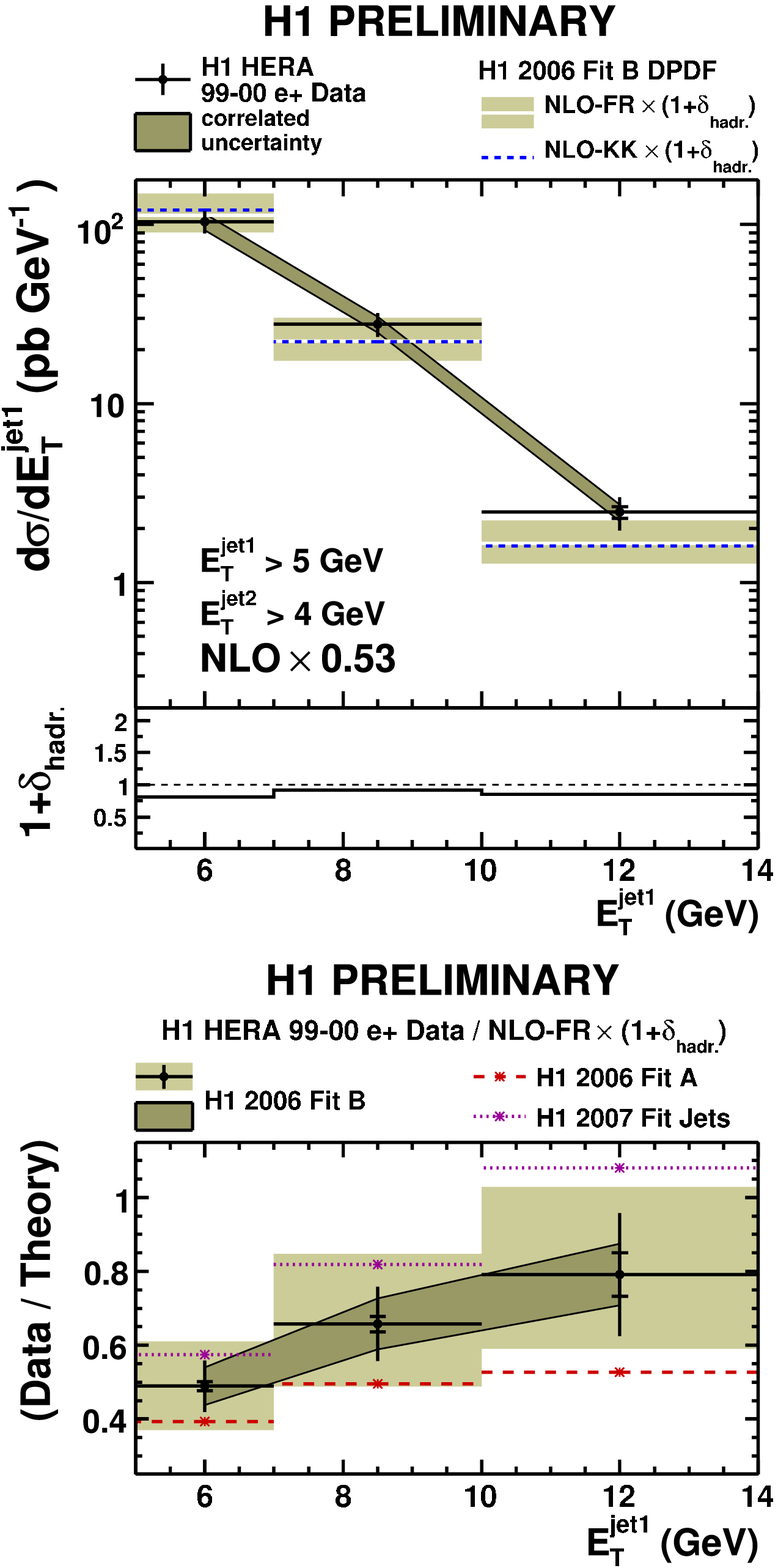}}
\caption{Distribution of the differential cross section $d\sigma/dE_T^{jet1}$
  normalised to the NLO-FR calculation based on the H1 2006 Fit B DPDF set and
  corrected for hadronisation effects. The points show the data, the inner
  error bars on the points are statistical and the outer error bars show
  statistical and uncorrelated uncertainties added in quadrature. The
  correlated systematic errors are indicated by the dark band. The effect on
  the NLO-FR calculation of varying $\mu_r$ by factors of $0.5$ and $2.0$ is shown
  in the light band. Also indicated are the central values obtained when the
  H1 2006 Fit A or H1 2007 Fit Jets DPDFs are used.
}
\label{fig:diffdijets}
\end{figure}

The H1 collaboration has performed a new measurement of the diffractive 
photoproduction of dijets \cite{ref:talk_diffjet}. They are performed in two different
kinematic regions primarily differing in the minimum transverse energy cuts 
of $5$ $(7.5)$~GeV of the two hardest jets. 
The cross sections are compared with QCD NLO
predictions \cite{ref:FR}, which overshoot the data in both cases. 
The ratio of the measured cross section over 
the QCD NLO prediction is shown in
Fig.~\ref{fig:diffdijets} for the case $E_T^{jet1}>5$~GeV as function of $E_T^{jet1}$.
These new results show a suppression as function of $E_T^{jet1}$ 
and suggest a suppression factor of about $0.5$ at small values of $E_T^{jet1}$.
The suppression is found to have no significant dependence on the photon
four-momentum fraction entering the hard subprocess and confirms previous
results obtained by the H1 collaboration \cite{ref:pub_diffjet}. 

\subsection{Physics with Flavours}
The H1 collaboration reported in \cite{ref:pub_charmpq} evidence for a narrow
resonance decaying into a $D^{\pm *}$ meson and a proton at an invariant mass of
$3.1$~GeV. The charmed meson was identified in the golden decay channel
$D^{\pm*} \rightarrow K^\mp\pi^\pm\pi^\pm$ and low momentum protons were
identified by means of measuring the specific energy loss.
This $D^*p$ resonance was not seen by other collider experiments and
could also not be confirmed by the ZEUS collaboration.
Using the full HERA II statistics this analysis is repeated in a limited
phase space region due to the smaller acceptance of the backward SPACAL
detector at HERA II. This new sample corresponds to an integrated luminosity of
${\cal L}=348$pb$^{-1}$, which is about 5 times the luminosity of the previously
analysed HERA~I data, where the excess was observed. 
For the new HERA~II data no excess of events in the region of
interest is observed, see  \cite{ref:talk_charmpq}. An upper limit on the
production of such a resonance is set at $1.0$ per mille on the ratio of
resonant $D^*p$ to $D^*$ production at 95\% confidence level using statistical errors only.

\begin{figure}[tp]
\vskip -0.7cm
\centerline{
\includegraphics[width=0.62\columnwidth]{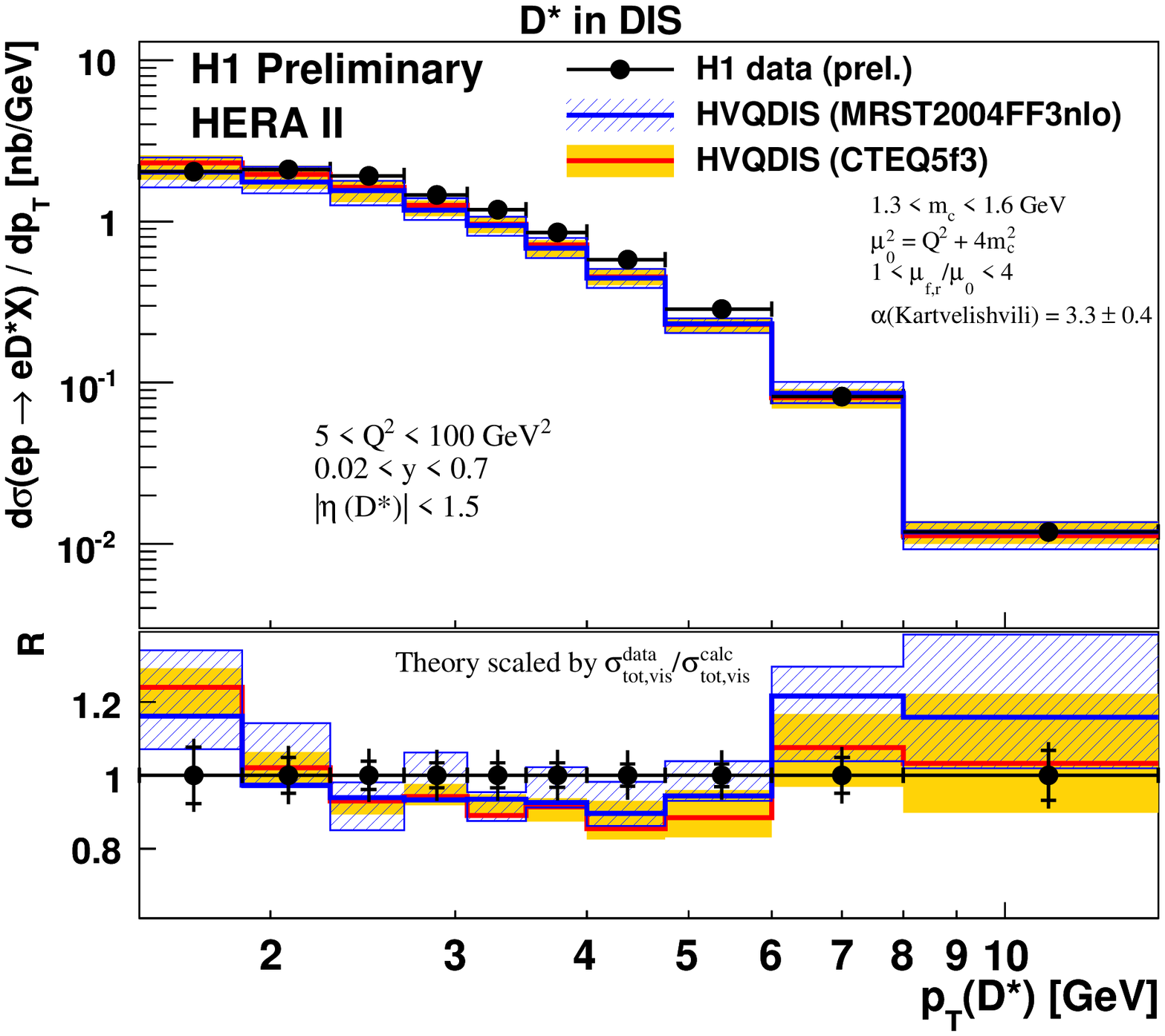}}
\vskip -0.2cm
\caption{Differential DIS cross section of $D^*$ mesons as a function of 
$p_T(D^*)$.
The measurements are given as black points; the inner error bars correspond to
the statistical, the outer to the total systematic error. The bands
reflecting the theoretical uncertainties
are obtained by varying the charm mass, the scale and the fragmentation parameters.
In the lower part the ratio of the normalised
differential cross sections data over prediction is shown.
}
\label{fig:dstarpt}
\centerline{
\includegraphics[width=0.62\columnwidth]{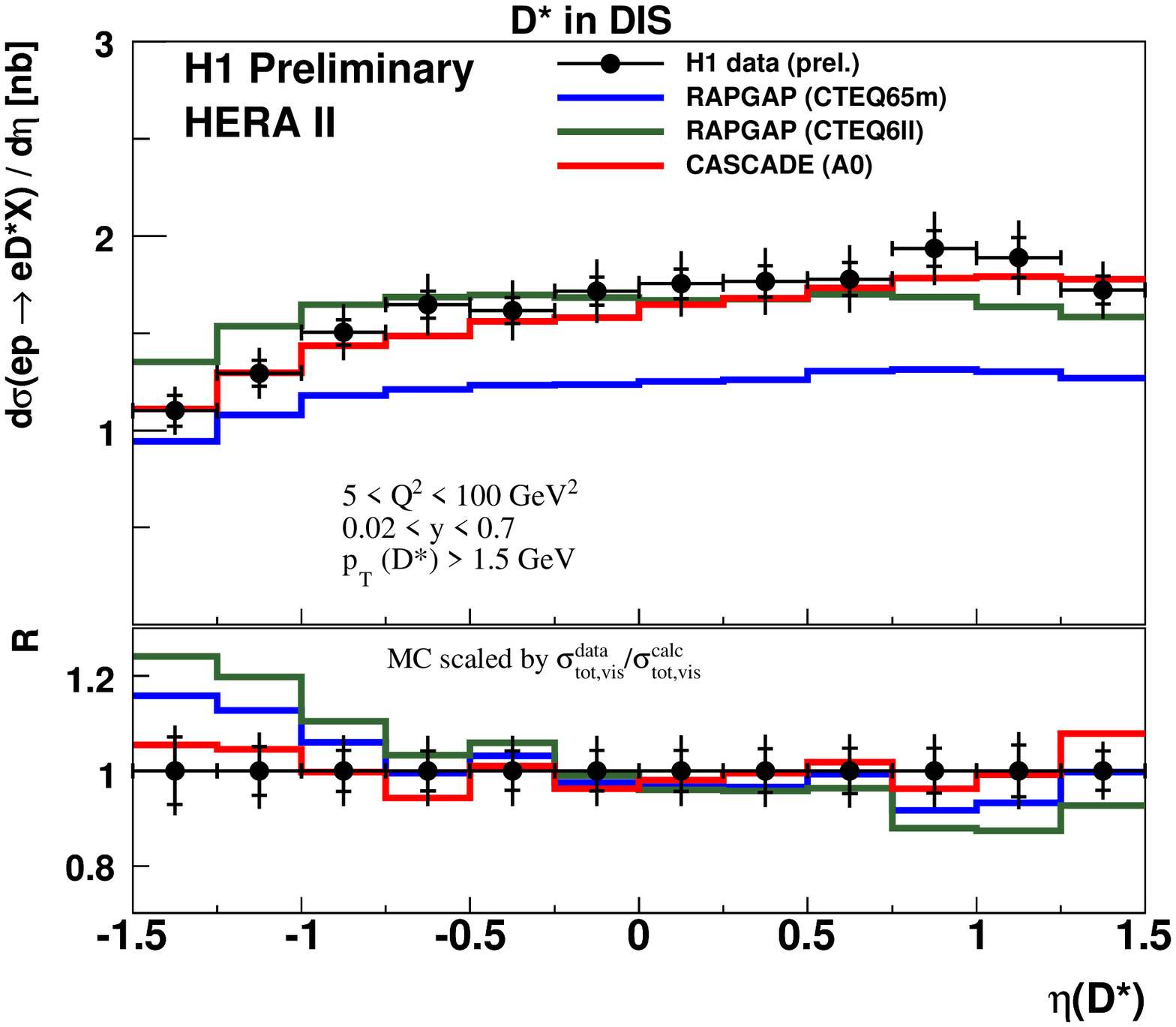}}
\vskip -0.2cm
\caption{Differential DIS cross section of $D^*$ mesons
as a function of the pseudorapidity.
For comparison LO Monte Carlo predictions from RAPGAP with PDFs from CTEQ65m (blue
histogram) and from CTEQ6ll (green histogram) and from CASCADE with the A0
unintegrated PDF (red histogram) are shown.
See also Fig.~\ref{fig:dstarpt} for explanations.
}
\label{fig:dstareta}
\end{figure}

New results on the study of $D^*$ production in DIS 
using the full HERA~II dataset are presented on this conference, see also 
\cite{ref:talk_dstar}. 
The data were taken in the years 2004-2007 and
correspond to an integrated luminosity of ${\cal L}=347$~pb$^{-1}$. 
Single and double differential cross sections are measured in the kinematic
range defined by $5<Q^2<100$~GeV$^2$ and $0.02<y<0.7$. The visible range of
the $D^*$ mesons is defined by $p_T(D^*)>1.5$~GeV and $|\eta(D^*)|<1.5$.
The results are compared with the LO Monte Carlo programs RAPGAP and CASCADE, and
with the NLO program HVQDIS. 
The distribution of the transverse momentum of the selected $D^*$ mesons $p_T(D^*)$ is shown
in Fig.~\ref{fig:dstarpt} and compared with the NLO predictions from HVQDIS for two
different proton PDFs. Within the errors agreement with the NLO calculation and
the MRST2004FF3nlo set \cite{ref:MRST2004FF3nlo} of the proton densities is
found. The data agree not so well with the prediction based on the CTEQ5f3 
PDF set \cite{ref:CTEQ5f3}.
The largest contribution to the uncertainties is of theoretical nature and comes from the scale uncertainty.

The sensitivity to the PDFs is also visible in the pseudorapidity distribution, 
see Fig.~\ref{fig:dstareta}, where the measured data points are
compared to LO Monte Carlo programs. The best description is found for
CASCADE, which is
based on unintegrated parton densities. However, it has to be noted
that the prediction also depends on the choice of the charm quark mass and 
the scale parameters.

\begin{wrapfigure}{r}{0.6\columnwidth}
\vskip -0.5cm
\centerline{
\includegraphics[width=0.62\columnwidth]{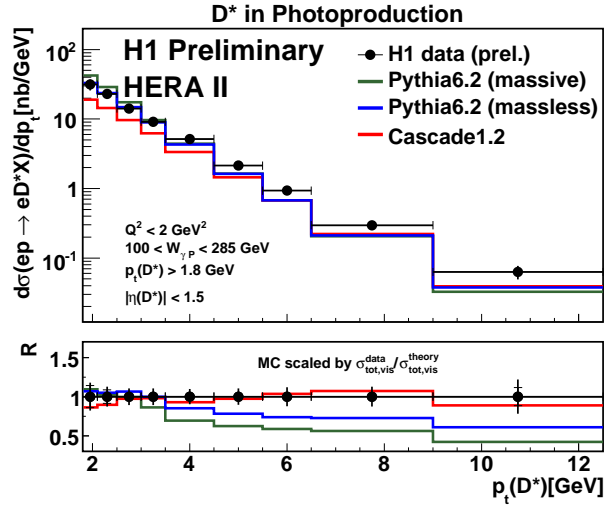}}
\vskip -2.5cm
\caption{Differential cross section of photoproduced $D^*$ mesons as a function of $p_T(D^*)$. 
The result is compared to different LO-MC models
The green histogram shows the prediction of the PYTHIA62 in the massive
scheme, whereas the blue histogram shows the prediction by PYTHIA62 is the
massless scheme. The red histogram shows the prediction of CASCADE12.
See also Fig.~\ref{fig:dstarpt} for explanations.
}
\label{fig:dstargp}
\end{wrapfigure}
The production of $D^*$-mesons can also be studied in the photoproduction
regime. A new analysis is presented based on data taken with a dedicated track trigger in
the years 2006-2007 corresponding to an integrated luminosity of ${\cal
  L}=93$~pb$^{-1}$.
$D^*$-mesons with a minimum transverse momentum of $p_T(D^*)>1.8$~GeV are selected 
in the pseudorapidity range 
$|\eta(D^*)|<1.5$. Fig.~\ref{fig:dstargp} shows the $p_T(D^*)$ distribution
of this measurement
and the comparison with different LO MC programs. 
Good agreement of the data with the massless PYTHIA62 prediction is found
at low $p_T(D^*)$. At high $p_T(D^*)$
all predictions underestimate the cross section. The shape of the distribution
is best described by the CASCADE program, which on the other hand largely
underestimates the absolute cross section. More detailed studies are presented
in \cite{ref:talk_dstar}, also including comparisons to NLO predictions, which
however suffer large scale uncertainties.

\section{Summary}
A selection of recent results obtained by the H1 collaboration was presented.
The complete list of all new H1 results can be found here \cite{url}.
One of the highlights is 
the first measurement of the structure
function $F_L$ at low $x$, which is based on dedicated 
datasets taken in 2007 with reduced proton beam energies. These measurements are
performed in a large range of $Q^2$ and are found to be consistent with pQCD predictions.

Many new results like the measurements of the DIS cross sections 
or the determination of $\alpha_S$ have reached high
precision by improving the understanding of the detectors, by
exploiting the full HERA datasets and by combination of the H1 and ZEUS data.
The HERA experiments are entering the era of high precision measurements
with full HERA statistics.  Many more exciting results are expected to come
in the near future.


\begin{footnotesize}

\end{footnotesize}


\end{document}